\input epsf
%%%%%%%%%%%%%%%%%%%%%%%%%%%%%%%%%%%%%%%%%%%%%%%%%%%%%%%%%%%%%%%%%
%								%
%	FONT FAMILIES:						%
%								%
%%%%%%%%%%%%%%%%%%%%%%%%%%%%%%%%%%%%%%%%%%%%%%%%%%%%%%%%%%%%%%%%%
%								%
%	Define script letters as rsfs		 		%
%		(or redefine as cal)				%
%		 						%
% 								%
%%%%%%%%%%%%%%%%%%%%%%%%%%%%%%%%%%%%%%%%%%%%%%%%%%%%%%%%%%%%%%%%%
\newfam\scrfam
\batchmode\font\tenscr=rsfs10 \errorstopmode
\ifx\tenscr\nullfont
	\message{rsfs script font not available. Replacing with calligraphic.}
\else	\font\sevenscr=rsfs7 
	\font\fivescr=rsfs5 
	\skewchar\tenscr='177 \skewchar\sevenscr='177 \skewchar\fivescr='177
	\textfont\scrfam=\tenscr \scriptfont\scrfam=\sevenscr
	\scriptscriptfont\scrfam=\fivescr
	\def\scr{\fam\scrfam}
	\def\cal{\scr}
\fi
%%%%%%%%%%%%%%%%%%%%%%%%%%%%%%%%%%%%%%%%%%%%%%%%%%%%%%%%%%%%%%%%%
%								%
%	Blackboard bold (or redefine as boldface)		%
%								%
%%%%%%%%%%%%%%%%%%%%%%%%%%%%%%%%%%%%%%%%%%%%%%%%%%%%%%%%%%%%%%%%%
\newfam\msbfam
\batchmode\font\twelvemsb=msbm10 scaled\magstep1 \errorstopmode
\ifx\twelvemsb\nullfont\def\Bbb{\bf}
	\message{Blackboard bold not available. Replacing with boldface.}
\else	\catcode`\@=11
	\font\tenmsb=msbm10 \font\sevenmsb=msbm7 \font\fivemsb=msbm5
	\textfont\msbfam=\tenmsb
	\scriptfont\msbfam=\sevenmsb \scriptscriptfont\msbfam=\fivemsb
	\def\Bbb{\relax\expandafter\Bbb@}
	\def\Bbb@#1{{\Bbb@@{#1}}}
	\def\Bbb@@#1{\fam\msbfam\relax#1}
	\catcode`\@=\active
\fi
%%%%%%%%%%%%%%%%%%%%%%%%%%%%%%%%%%%%%%%%%%%%%%%%%%%%%%%%%%%%%%%%%
%								%
%	MORE FONTS:						%
%								%
%%%%%%%%%%%%%%%%%%%%%%%%%%%%%%%%%%%%%%%%%%%%%%%%%%%%%%%%%%%%%%%%%
\font\eightrm=cmr8		\def\xrm{\eightrm}
\font\eightbf=cmbx8		\def\xbf{\eightbf}
\font\eightit=cmti10 at 8pt	\def\xit{\eightit}
\font\eighttt=cmtt8		\def\xtt{\eighttt}
\font\eightcp=cmcsc8
\font\eighti=cmmi8		\def\xold{\eighti}
\font\teni=cmmi10		\def\old{\teni}

\font\twelverm=cmr12
\font\twelvecp=cmcsc10 scaled\magstep1
\font\fourteencp=cmcsc10 scaled\magstep2
\font\fiverm=cmr5
\def\ss{\scriptstyle}
\def\sss{\scriptscriptstyle}
%%%%%%%%%%%%%%%%%%%%%%%%%%%%%%%%%%%%%%%%%%%%%%%%%%%%%%%%%%%%%%%%%
%								%
%	HEADLINE:						%
%								%
%%%%%%%%%%%%%%%%%%%%%%%%%%%%%%%%%%%%%%%%%%%%%%%%%%%%%%%%%%%%%%%%%
\headline={\ifnum\pageno=1\hfill\else
{\eightcp Cederwall, von Gussich, Mikovi\'c, Nilsson, Westerberg: 
	``On the DBI Action for D-Branes''}
		\dotfill{ }{\old\folio}\fi}
\def\makeheadline{\vbox to 0pt{\vss\noindent\the\headline\break
\hbox to\hsize{\hfill}}
	\vskip2\baselineskip}
%%%%%%%%%%%%%%%%%%%%%%%%%%%%%%%%%%%%%%%%%%%%%%%%%%%%%%%%%%%%%%%%%
%								%
%       FOOTNOTES:						%
%								%
%%%%%%%%%%%%%%%%%%%%%%%%%%%%%%%%%%%%%%%%%%%%%%%%%%%%%%%%%%%%%%%%%
%\footline={\ifnum\pageno=1\else\hfil\folio\hfil\fi}
\def\makefootline{\ifnum\foottest=1
	\baselineskip=.8cm\line{\the\footline}\global\foottest=0
        %\else\baselineskip=1.6cm\line{}
	\fi
        }
\newcount\foottest
\foottest=0
\def\footnote#1#2{${\,}^#1$\footline={\vtop{\baselineskip=9pt
        \hrule width.5\hsize\hfill\break
        \indent ${}^#1$ \vtop{\hsize=14cm\noindent\xrm #2}}}\foottest=1
        }
%%%%%%%%%%%%%%%%%%%%%%%%%%%%%%%%%%%%%%%%%%%%%%%%%%%%%%%%%%%%%%%%%
%								%
%	REFERENCES:						%
%								%
%%%%%%%%%%%%%%%%%%%%%%%%%%%%%%%%%%%%%%%%%%%%%%%%%%%%%%%%%%%%%%%%%
\newcount\refcount
\refcount=0
\newwrite\refwrite
\def\ref#1#2{\global\advance\refcount by 1
	\xdef#1{{\old\the\refcount}}
	\ifnum\the\refcount=1
	\immediate\openout\refwrite=\jobname.refs
	\fi
	\immediate\write\refwrite
		{\item{[{\xold\the\refcount}]} #2\hfill\par\vskip-2pt}}
\def\refout{\catcode`\@=11 
	\xrm\immediate\closeout\refwrite
	\vskip\baselineskip
	{\noindent\twelvecp References}\hfill%\vskip\baselineskip
						\vskip.25\baselineskip%%%%
	\parskip=.875\parskip 
	\baselineskip=.8\baselineskip
	\input\jobname.refs 
	\parskip=8\parskip \divide\parskip by 7
	\baselineskip=1.25\baselineskip 
	\catcode`\@=\active\rm}
%%%%%%%%%%%%%%%%%%%%%%%%%%%%%%%%%%%%%%%%%%%%%%%%%%%%%%%%%%%%%%%%%
%								%
%	EQUATION NUMBERING					%
%								%
%%%%%%%%%%%%%%%%%%%%%%%%%%%%%%%%%%%%%%%%%%%%%%%%%%%%%%%%%%%%%%%%%
\newcount\eqcount
\eqcount=0
\def\Eqn#1{\global\advance\eqcount by 1
	\xdef#1{%\the\sectioncount.
	{\old\the\eqcount}}
		\eqno(%\the\sectioncount.
		{\oldstyle\the\eqcount})}
\def\eqn{\global\advance\eqcount by 1
	\eqno(%\the\sectioncount.
	{\oldstyle\the\eqcount})}
\def\multi{\global\advance\eqcount by 1}
\def\multieq#1#2{\xdef#1{{\old\the\eqcount#2}}
	\eqno{({\oldstyle\the\eqcount#2})}}
%%%%%%%%%%%%%%%%%%%%%%%%%%%%%%%%%%%%%%%%%%%%%%%%%%%%%%%%%%%%%%%%%	
%								%
%	FORMAT:							%
%								%
%%%%%%%%%%%%%%%%%%%%%%%%%%%%%%%%%%%%%%%%%%%%%%%%%%%%%%%%%%%%%%%%%
\parskip=3.5pt plus .3pt minus .3pt
\baselineskip=12.5pt plus .1pt minus .05pt
\lineskip=.5pt plus .05pt minus .05pt
\lineskiplimit=.5pt
\abovedisplayskip=10pt plus 4pt minus 2pt
\belowdisplayskip=\abovedisplayskip
\hsize=15cm
\vsize=20cm
\hoffset=1cm
\voffset=1.3cm
%%%%%%%%%%%%%%%%%%%%%%%%%%%%%%%%%%%%%%%%%%%%%%%%%%%%%%%%%%%%%%%%%
%								%
%	VARIOUS DEFINITIONS					%
%								%
%%%%%%%%%%%%%%%%%%%%%%%%%%%%%%%%%%%%%%%%%%%%%%%%%%%%%%%%%%%%%%%%%
\def\/{\over}
\def\*{\partial}
\def\a{\alpha}
\def\b{\beta}
\def\e{\varepsilon}
\def\fh{\hat{F}}
\def\g{\gamma}
\def\gg{\g^{-1}}
\def\k{\kappa}
\def\l{\lambda}
\def\E{{\Bbb E}}
\def\B{{\Bbb B}}
\def\D{\Delta}

\def\L{{\cal L}}
\def\punkt{\,\,.}
\def\komma{\,\,,}
\def\.{.\hskip-1pt }
\def\is{\!=\!}
\def\-{\!-\!}
\def\+{\!+\!}
\def\={\!=\!}
\def\>{\!>\!}
\def\half{{1\/2}}
\def\quarter{{1\/4}}
\def\tr{\hbox{tr}}
\def\ie{{\it i.e.}\hskip-1pt}
\def\eg{{\it e.g.}\hskip-1pt}
\def\DBI{\hbox{Dirac--Born--Infeld}}
\def\dbi{{\hbox{\fiverm DBI}}}
\def\det{\hbox{\rm det}}
\def\fun{\varphi}
\def\II{I\hskip-0.6pt I}

%%%%%%%%%%%%%%%%%%%%%%%%%%%%%%%%%%%%%%%%%%%%%%%%%%%%%%%%%%%%%%%%%%
%%%%%%%%%%%%%%%%%%%%%%%%%%%%%%%%%%%%%%%%%%%%%%%%%%%%%%%%%%%%%%%%%%
%%								%%
%%	THE PAPER						%%
%%								%%
%%%%%%%%%%%%%%%%%%%%%%%%%%%%%%%%%%%%%%%%%%%%%%%%%%%%%%%%%%%%%%%%%%
%%%%%%%%%%%%%%%%%%%%%%%%%%%%%%%%%%%%%%%%%%%%%%%%%%%%%%%%%%%%%%%%%%

\null\vskip-2cm
\hbox to\hsize{\hfill G\"oteborg-ITP-{\old96}-{\old8}}
\hbox to\hsize{\hfill\tt hep-th/9606173}
\hbox to\hsize{\hfill June, {\old1996}}

\epsfxsize=9cm
\vskip1.5cm\hskip2.2cm
\epsffile{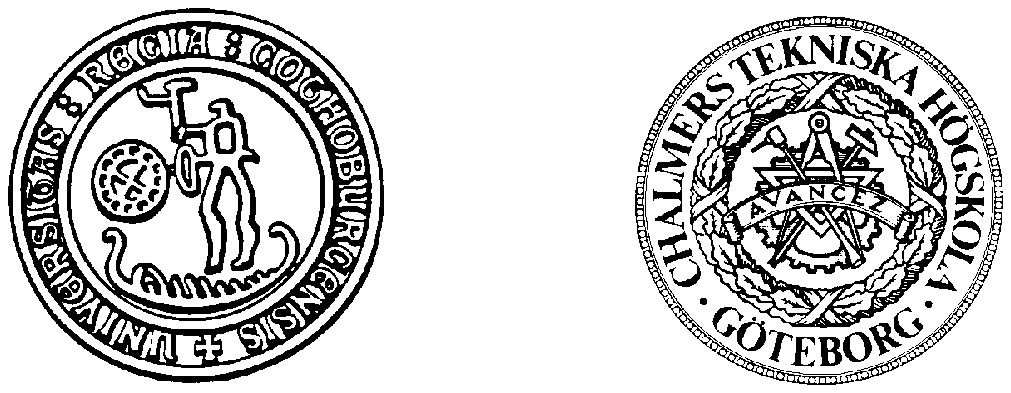}

\vskip2cm
\centerline{\fourteencp On the Dirac--Born--Infeld Action for D-Branes} 
\vskip\parskip
\centerline{\twelvecp}

\vskip1.5cm
\centerline{\twelverm Martin Cederwall, Alexander von Gussich,}
\vskip8pt
\centerline{\twelverm Aleksandar Mikovi\'c${}^*$, 
      Bengt E.W. Nilsson and Anders Westerberg}

\vskip1cm
\centerline{\it Institute of Theoretical Physics}
\centerline{\it G\"oteborg University and Chalmers University of Technology }
\centerline{\it S-412 96 G\"oteborg, Sweden}

\vskip1cm

\noindent \underbar{\it Abstract:}{ } 
In this note, we consider the reformulation of 
the Dirac--Born--Infeld action for a Dirichlet $p$-brane in 
Brink--Di~Vecchia--Howe--Tucker form, \ie, including an
independent non-propagating world-volume metric.
When $p\>2$, the action becomes non-polynomial.
A closed expression is derived for $p\is3$. For selfdual field-strengths,
the DBI action is reproduced by an action with a simple $F^2$ term.
We speculate on supersymmetrization of the D${}_3$-brane action.
We also give the governing equations for arbitrary $p$, and derive an
implicit expression for the D${}_4$-brane lagrangian.
  
\vfill

\catcode`\@=11
\vtop{\baselineskip=9pt
\hbox to\hsize{\xrm email addresses: \xtt tfemc@fy.chalmers.se\hfill}
\hbox to\hsize{\xrm \phantom{email addresses: }\xtt 
	tfeavg@fy.chalmers.se\hfill}
\hbox to\hsize{\xrm \phantom{email addresses: }\xtt mikovic@fy.chalmers.se,
     	mikovic@castor.phy.bg.ac.yu\hfill}
\hbox to\hsize{\xrm \phantom{email addresses: }\xtt tfebn@fy.chalmers.se\hfill}
\hbox to\hsize{\xrm \phantom{email addresses: }\xtt 
	tfeawg@fy.chalmers.se\hfill}}
\catcode`\@=\active

\vskip10pt
\noindent ${}^*${\xrm{}Permanent address: Institute of Physics, 
	P.O. Box {\xold57}, Belgrade {\xold11001}, Yugoslavia.}

\eject

\def\nl{\hfill\break\indent}
\ref\Hull{C.M.~Hull and P.K.~Townsend, 
	{\xit ``Unity of Superstring Dualities''}, \nl Nucl.Phys. 
	{\xbf B438} ({\xold1995}) {\xold109} ({\xtt hep-th/9410167}).}
\ref\Witten{E.~Witten, {\xit ``String Theory Dynamics in Various Dimensions''},
	Nucl.Phys. {\xbf B443}
	({\xold1995}) {\xold85} ({\xtt hep-th/9503124}).}
\ref\Duff{M.J.~Duff, {\xit ``Strong/Weak Coupling Duality from the Dual
	String''}, Nucl.Phys. {\xbf B442}
	({\xold1995}) {\xold47} ({\xtt hep-th/9501030}).}
\ref\Khuri{M.J.~Duff, R.R.~Khuri and J.X.~Lu, {\xit ``String Solitons''},
	Phys.Rep. {\xbf 259} ({\xold1995}) {\xold213} ({\xtt hep-th/9412184}).}
\ref\Dai{J.~Dai, R.G.~Leigh and J.~Polchinski, 
	{\xit ``New Connections between String Theories''}, 
	\nl Mod.Phys.Lett. {\xbf A4} ({\xold1989}) {\xold2073}.}   
\ref\Leigh{R.G. Leigh, {\xit ``Dirac--Born--Infeld Action from Dirichlet Sigma
	Model''}, Mod.Phys.Lett.~{\xbf A4} ({\xold1989}) {\xold2767}.}
\ref\Polchinski{J. Polchinski, {\xit ``Dirichlet-Branes and Ramond--Ramond 
	Charges''},
	\nl Phys.Rev.Lett.~{\xbf 75} ({\xold1995}) {\xold4724} 
	({\xtt hep-th/9510017}).}
\ref\Guven{R.~G\"uven, 
	{\xit ``Black p-Brane Solutions of D=11 Supergravity Theory''}, 
	Phys.Lett. {\xbf 276B} ({\xold1992}) {\xold49}.}
\ref\Townsend{P.K. Townsend, {\xit ``D-branes from M-branes''}, 
	Phys.Lett.~{\xbf 373B} ({\xold1996}) {\xold68} 
	({\xtt hep-th/9512062}).}
\ref\Schmidhuber{C. Schmidhuber, {\xit ``D-brane actions''}, 
	{\xtt hep-th/9601003}.}
\ref\Bergshoeff{E.~Bergshoeff, E.~Sezgin and P.K.~Townsend,
	{\xit ``Properties of the Elevendimensional Supermembrane Theory''}, 
	\nl Ann.Phys. {\xbf 185} ({\xold1988}) {\xold330}.}
\ref\Tucker{P.S.~Howe and R.W.~Tucker, 
	{\xit ``A Locally Supersymmetric and Reparametrization Invariant
	Action for a Spinning \nl Membrane''}, J.Phys. {\xbf A10}
		({\xold1977}) L{\xold155}.}
\ref\BDH{L.~Brink, P.~Di~Vecchia and P.~Howe, 
	{\xit ``A Locally Supersymmetric and Reparametrization Invariant 
	Action for the \nl Spinning String''},
	Phys.Lett. \xbf 65B \xrm ({\xold1976}) {\xold471}.}
\ref\DuffLu{M.J.~Duff and J.X.~Lu, 
	{\xit ``Type II p-branes: the brane-scan revisited''}, 
	\nl Nucl.Phys. {\xbf B390} ({\xold1993}) {\xold276} 
	({\xtt hep-th/9207060}).}
\ref\Douglas{M.~Douglas, {\xit ``Branes within Branes''},
	{\xtt hep-th/9512077}.}
\ref\Green{M.B.~Green, C.M.~Hull and P.K.~Townsend, \nl {\xit ``D-Brane 
	Wess--Zumino Actions, T-Duality and the Cosmological Constant''},
	{\xtt hep-th/9604119}.}
\ref\Tseytlin{A.A.~Tseytlin, {\xit ``Self-duality of Born-Infeld action and 
	Dirichlet 3-brane of type IIB superstring''},
	{\xtt hep-th/9602064}.}
\ref\JatkarRama{D.P.~Jatkar and S.K.~Rama, 
	{\xit ``F-theory from Dirichlet 3-branes''},
	{\xtt hep-th/9606009}.}
\ref\Blencowe{M.P.~Blencowe and M.J.~Duff, {\xit ``Supermembranes and the 
	Signature of Space-Time''}, \nl Nucl.Phys. {\xbf B310}
	({\xold1988}) {\xold387}.}
\ref\Vafa{C. Vafa, {\xit ``Evidence for F-Theory''}, {\xtt hep-th/9602022}.}
\ref\HullII{C.M.~Hull, {\xit ``String Dynamics at Strong Coupling''},
	{\xtt hep-th/9512181}.}

The introduction of $p$-branes in string theory has turned out to be 
instrumental in unraveling many of its non-perturbative properties
(see \eg\ [\Hull,\Witten,\Duff]). In type \II A and \II B string theory,
$p$-branes for various values of $p$ arise as solutions of the low-energy
field equations. They are of two kinds depending on whether the field equations
of which they are solutions involve antisymmetric tensor fields from the
NS-NS or the R-R sector. In the case of NS-NS fields, the $p$-branes are
interpreted as either fundamental or solitonic in the sense familiar from
ordinary field theory (see \eg\ [\Khuri,\Hull]). 
In the R-R sector, on the other
hand, the $p$-branes are neither fundamental nor genuinely solitonic, but 
have been found to have a description in terms of open strings
with mixed Dirichlet and Neumann boundary conditions, ending at the
location of the $p$-branes [\Dai,\Leigh,\Polchinski].
$p$-branes of this type are referred to as Dirichlet branes, or D-branes.
More recently, also the 5-brane solution [\Guven] of 11-dimensional 
supergravity has been given a similar interpretation in terms of open 
membranes~[\Townsend].

The fact that $p$-branes with R-R charges are located at the ends of open
strings implies that the $p$-brane world-volume field theory can be derived
from open string $\b$-functions [\Leigh]. For an abelian and constant 
field-strength $F_{mn}$, the result is a ``kinetic'' term of the
\DBI\ (DBI) type.
  
The action for the D${}_2$-brane is known to be equivalent 
[\Schmidhuber,\Townsend] to the 11-dimensional supermembrane [\Bergshoeff],
which is the world-volume metric formulation of Howe and Tucker [\Tucker] 
generalizing the Brink--Di~Vecchia--Howe string action [\BDH] to $p\!\geq\!2$
(we refer to the form of the action containing
the world-volume metric as the Brink--Di~Vecchia--Howe--Tucker (BDHT) form).
In three dimensions, a vector potential is
dual to a scalar, which can be interpreted as the eleventh coordinate 
[\DuffLu].
Therefore, also the $\k$-symmetric version is known 
[\DuffLu,\Bergshoeff,\Townsend].
For higher values of $p$, no exact BDHT-type actions for D-branes 
have been constructed so far, although 
the action to lowest order in derivatives is given in [\Townsend],
and only the bosonic parts of the DBI actions, including couplings to 
background fields, are known [\Douglas,\Green]. 
It is conceivable, and this is one of the main motivations for the present
work, that the construction of 
supersymmetric D${}_p$-brane actions will benefit from knowledge of a 
BDHT-type action.

The purpose of this note is to develop techniques that may be useful in
establishing links between the DBI forms of D-brane actions and their
corresponding BDHT forms. As emphasized by Townsend [\Townsend], these latter
forms are not likely to be simple for $p\>2$, but, rather, 
become non-polynomial
in the field-strength $F$. In fact, the highly non-linear equations arising
when varying with respect to the world-volume metric have not been solved
previously for $p\>1$ (although the $p\=2$ solution is implicit in 
[\Townsend]).
For higher values of $p$, not having general techniques for solving these
equations also means that the higher order terms in the BDHT formulation
can not be deduced. 

The methods developed here indicate that these problematic issues can in fact
be resolved. We will in particular demonstrate that, for $p\=3$, the DBI action
is equivalent to a BDHT action, which is non-polynomial, but of a reasonably
simple form. A remarkable simplification occurs if a self-duality constraint 
is imposed; the action then becomes bilinear in $F$. It is also interesting
to note that although the world-volume metric is not set equal to the induced
metric by its equation of motion, their determinants are equal, a fact that
may prove very useful. Other aspects of D${}_3$-brane actions have been 
addressed previously in [\Tseytlin,\JatkarRama].

Consider a general lagrangian for a $p$-brane with an
abelian vector potential\footnote{{\oldstyle1}}{We omit the dilaton factor
${\ss e}^{\sss -\phi}$,
which is irrelevant for our discussion. Also, couplings to R-R background 
fields are not considered, since they enter as Wess--Zumino terms, not 
containing the world-volume metric.}:
$$
\L=-\half\sqrt{-\g}\,\bigl\{\tr(\gg g)+\fun(\gg \fh)\-(p-1)\bigr\}
	\punkt\Eqn\generallagrangian
$$
Here $\g_{mn}$ is the world-volume metric, 
$g_{mn}=\*_mX^\mu\*_nX^\nu G_{\mu\nu}$ the induced metric 
(\ie, the pullback of the space-time metric) and 
$\fh_{mn}=F_{mn}-{2\pi\over{\a^\prime}}B_{mn}$ where $F_{mn}$ is 
the abelian field-strength and $B_{mn}$ the pull-back to the world-volume of
the NS-NS antisymmetric tensor field.
The function $\fun$ is some scalar function of $\gg \fh$, that should be 
chosen so that solving the equations of motion for $\g$ results in the
\DBI\ lagrangian
$$
\L_\dbi=-\sqrt{-\det(g+\fh)}\punkt\Eqn\DBIlagrangian
$$
As a warmup exercise, and in order to establish our procedure,
we first treat the case $p\is2$.
We define the matrices
$$
\eqalign{
&u=\gg g\komma\cr
&X=g^{-1}\fh\komma\cr}\Eqn\matrixdef
$$
and the objective is to use the equations of motion for the world-volume
metric in order to solve for $u$ in terms of $X$.
In three dimensions, the matrix $X$ obeys the identity
$$
X^3=\half X\tr X^2\punkt\eqn
$$
The only invariant to enter the lagrangian through $\fun$ 
is $\tr(uX)^2$, so we can set
$\fun=\fun(\tr(uX)^2)$. The equations of motion for $\gg$ become
$$
0=-{2\/\sqrt{-\g}}\gg{\*\L\,\,\/\*\gg}
	=-\half\tr u-\half\fun+\half+u+2(uX)^2\fun'
\punkt\Eqn\pistwoeom
$$
It is clear that $u$ only contains even powers of $X$, and a convenient
basis turns out to be $u=A\+B(X^2\-\half\tr X^2)$, 
where $A$ and $B$ are functions of $\tr X^2$. This turns the equations of 
motion (\pistwoeom) into
$$
0=\half-\half A-{1\/4}B\tr X^2-\half\fun(A^2\tr X^2)
	+X^2\,\bigl[\,2A^2\fun'(A^2\tr X^2)+B\,\bigr]\punkt\Eqn\ptwocompeom
$$
When $\fun$ is just the ordinary $\fh^2$ term, $\fun(t)\is-\half t$,
the solution is $A=B=1$. We insert the solution back into $\L$,
using $\det u=\det(1\+X^2\-\half\tr X^2)=1\-\half\tr X^2=\det(1\+X)$
to obtain
$$
\eqalign{
\L=&-\sqrt{-\g}\,\bigl(1-\half\tr X^2\bigr)
		=-\sqrt{-g}\,(\det u)^{-1/2}\bigl(1-\half\tr X^2\bigr)\cr
	=&-\sqrt{-g\phantom{X}\hskip-10pt}\,\sqrt{\det(1+X)}
		=-\sqrt{-\det(g+\fh)}\komma\cr}\eqn
$$
and the equivalence to the \DBI\ action is established.

We now move on to $p\is3$. The relevant matrix identity is
$$
X^4=\half X^2\tr X^2-\det X\punkt\Eqn\pthreeidentity
$$
The function $\fun$ will depend on the invariants $\tr(uX)^2$ and 
$\det(uX)$, and $u$ can again be expanded as, \eg, $u\=A\+BX^2$.
It turns out to be more convenient, however, to choose another basis, 
namely one where $X^2$ acts diagonally. From (\pthreeidentity) it follows
that 
$$
X^2v_\pm=\l_\pm v_\pm\komma\eqn
$$
where
$$
\eqalign{
&\l_\pm={1\/4}\tr X^2\pm\sqrt{-\D}\komma\cr
&v_\pm=\half\left[\,1\pm{1\/\sqrt{-\D}}\Bigl(X^2-{1\/4}\tr X^2\Bigr)\right]
	\punkt\cr}\Eqn\eigenthings
$$
The scalar $\D$, that will be of some interest later, is defined as
$\D(X)=\det X\!-\!{1\/16}(\tr X^2)^2$.
It is negative semi-definite. The matrices $v_\pm$ are normalized to 
become projection operators on the subspaces with the associated 
eigenvalues. We now expand $u$ as $u=u_+v_++u_-v_-$, and obtain
$$
\eqalign{
&\tr(uX)^2=2\,(\l^{\phantom{2}}_+u_+^2+\l^{\phantom{2}}_-u_-^2)\komma\cr
&\det(uX)=\l^{\phantom{2}}_+\l^{\phantom{2}}_-u_+^2u_-^2\komma\cr}\eqn
$$
so that the function $\fun$ in the lagrangian (\generallagrangian) is
a function of $\l^{\phantom{2}}_+u_+^2$ and $\l^{\phantom{2}}_-u_-^2$. 
After some manipulations, the
equations of motion for $\gg$ become
$$
\eqalign{
&1-\half\fun-u_-+\half u_+{\*\fun\,\,\,\/\*u_+}=0\komma\cr
&1-\half\fun-u_++\half u_-{\*\fun\,\,\,\/\*u_-}=0\punkt\cr}
	\Eqn\pthreeeom
$$
The occurrence of derivatives should not be interpreted as if these 
equations were differential equations --- for a given function $\fun$ they
are simply algebraic equations for $u_\pm$. If, in addition, we want the 
action to be equivalent to the \DBI\ action, we use 
$\det(1\+X)\=(1\-\l_+)(1\-\l_-)$ to obtain the condition 
$$
1-u_+-u_--\half\fun+u_+u_-\sqrt{(1\-\l_+)(1\-\l_-)}=0\punkt\Eqn\DBIequivalence
$$
The resulting set of equations is seemingly difficult to solve. After 
solving for $\fun$ and $u_\pm$ order by order a couple of steps, we noticed
that we got $\det u\equiv u_+^2u_-^2=1$, \ie, $\det\g\=\det g$. 
We do not understand why this happens,
but by assuming it to hold exactly one can considerably restrict the 
possible functions $\fun$ by demanding them to give solutions with  
$\det u\=1$ (this will be verified by the exact solution). The search is
simplified by the change of variables to 
$s_\pm\=(2\l^{\phantom{2}}_\pm u_\pm^2)^{-1}$,
and $\chi=2(s_+s_-)^{1/2}(1\-\half\fun)$. 
The determinant condition then becomes
$$
{\*\chi\,\,\,\/\*s_+}{\*\chi\,\,\,\/\*s_-}=1\komma\Eqn\determinantone
$$
and the equations of motion (\pthreeeom) together with the condition
(\DBIequivalence) turn into
\multi
$$
{\*\chi\,\,\/\*s_+}-{1\/\sqrt{2\l_-s_+}}=0\komma\multieq{\spmeom}{a}
$$
$$
\sqrt{\l_+\l_-}\,\chi-{1\/\sqrt{2\l_+s_+}}
	-\sqrt{2\l_+s_+}+\sqrt{(1\-\l_+)(1\-\l_-)}=0\punkt
		\multieq{\spmdetrel}{b}
$$
It is not difficult to find a number of functions satisfying 
(\determinantone),
and if the system is going to be soluble, we can not afford complicated
functions leading to transcendental equations. The simplest reasonable
function is $\chi\is2(s_+s_-)^{1/2}$, which of course corresponds to 
$\fun\is0$. We can modify it to $\chi\is2((s_+\+\a)(s_-\+\a))^{1/2}$, and
it turns out that the solution of (\spmeom), 
$$
\eqalign{
&s_+={1\/2\l_+}{1-2\a\l_+\/1-2\a\l_-}\komma\cr
&s_-={1\/2\l_-}{1-2\a\l_-\/1-2\a\l_+}\komma\cr}\eqn
$$
fulfills (\spmdetrel) for $\a\is\half$. 
This gives the solution for the 
BDHT-type D${}_3$-brane lagrangian. Expressed in terms of the 
familiar variables
$\tr(uX)^2$ and $\det(uX)$ the function $\fun$ is
$$
\eqalign{
\fun\,=\,&\,2\,\biggl\{1-\sqrt{\bigl(1+\l_+^{\phantom{2}}u_+^2\bigr)
		\bigl(1+\l_-^{\phantom{2}}u_-^2\bigr)}\,\,\biggr\}\cr
	\,=\,&\,2\,\biggl\{1-\sqrt{1+\half\tr(uX)^2+\det(uX)}\,\,\biggr\}\cr
	\,=\,&\,2\,\biggl\{1
		-\sqrt{\bigl(1+{1\/4}\tr(uX)^2\bigr)^2+\D(uX)}\,\,\biggr\}\cr
	\,=\,&\,2\,\biggl\{1-\Bigl[\det\bigl(1+(uX)^2\bigr)\Bigr]^{1/4}
		\,\biggr\}
		\komma\cr}\eqn
$$
so the lagrangian reads
$$
\L=\sqrt{-\g}\,\biggl\{\,-\half\tr(\gg g)
		+\sqrt{\bigl(1+{1\/4}\tr(\gg \fh)^2\bigr)^2+\D(\gg \fh)}
		\,\,\biggr\}
          \Eqn\funnallagrangian
$$
(note that the cosmological constant gets absorbed in the square root).
The expression inside the square root is never negative.
The occurrence of a square root should not be associated to the square
root in the DBI action, it is rather a feature particular to $p\=3$.

Notice that if $\D(\gg \fh)$ vanishes, the lagrangian reduces to one containing
only the $\fh^2$ term, as for $p\is2$. When we examine the meaning of this
condition, we have to be careful about the signature of the world-volume
metric. If it is lorentzian, as for the application we set out to
investigate, both terms in $\D$ are negative semi-definite, so here it 
implies that $\det(\gg \fh)\=0\=\tr(\gg \fh)^2$, or equivalently,
$\E\!\cdot\!\B\=0\=\E^2\-\B^2$. Then all terms in $\L$ containing $\fh$ vanish,
and this case becomes trivial. 
If, on the other hand, the world-volume signature
is $(4,0)$ or $(2,2)$, the condition is equivalent to selfduality or
anti-selfduality (with respect to the world-volume metric). 
When we analyze this action, with a selfduality constraint, it turns out
that it is irrelevant which metric (world-volume or induced) we refer to
when selfduality is imposed. We can not use the basis (\eigenthings), which
is not well defined when $\D\is0$, so we write $u\=A\+B(X^2\-{1\/4}\tr X^2)$,
and the solution is straightforward, yielding $A\=1$, 
$B\=(1\-\half\tr X^2)^{-1}$. The \DBI\ lagrangian is again reproduced.
We believe that this case may be relevant for the formulation of F-theory
(see \eg\ [\Blencowe,\Vafa,\HullII]).

There are a couple of questions that deserve further investigation.
One is the generalization to higher $p$. It is not clear to us whether
it will be possible to obtain closed forms for the BDHT-type
lagrangians for $p\>3$. 
It is conceivable that the equations simply
become too complicated, but on the other hand, this was what we thought
would happen already for $p\is3$.
 
The generalization of the method we have used for $p\is3$ is straightforward,
and probably gives the simplest formulation of the problem.
For the case of odd $p$, $p\is2n\-1$, there are $n$ non-zero eigenvalues $\l_i$
for $X^2$, each with an associated subspace of dimension two,
$\tr v_i\is2$. The equations of motion for $\gg$ together with the
condition that the action reduces to the \DBI\ action read
$$
\eqalign{
	&0=n\-1-\sum_{j\neq i}u_j-\half\fun+\half u_i{\*\fun\/\*u_i}\komma
		\quad i\=1,\ldots,n,\cr
	&\sum_iu_i+\half\fun-(n\-1)=\prod_i(1\-\l_i)^{1/2}u_i\punkt\cr}
	\eqn
$$
For the case of even $p$, $p\is2n$, there are $n$ non-zero eigenvalues $\l_i$
for $X^2$, each with an associated linear subspace of dimension two,
$\tr v_i\is2$, and one zero eigenvalue with $\tr v_0\is1$. The $v_0$ component
of $u$ does not enter in $\fun$, and can be solved for,
$$
u_0=-(2n\-1)+2\sum_iu_i+\fun\punkt\eqn
$$
The remaining equations read
$$
\eqalign{
	&0=-u_0+u_i+\half u_i{\*\fun\/\*u_i}\komma\quad i\=1,\ldots,n,\cr
	&u_0=\prod_i(1\-\l_i)u_i^2\punkt\cr}\eqn
$$
It seems difficult in general to determine $\fun$ explicitly.
For $p\is4$ we have been able to find the explicit solution for $u$.
Here we have two non-zero eigenvalues that we denote $\l_\pm$, which expressed
in terms of $X\is\gg\fh$ are 
$\l_\pm=\quarter\tr X^2\pm
[\,\quarter\tr X^4\-{1\/16}(\tr X^2)^2\,]^{1/2}$.
The solution is 
$$
u_\pm=(1\-\l_\pm)^{-2/3}\,(1\-\l_\mp)^{1/3}\punkt\eqn
$$
The function $\fun$, which is a function of $t_\pm\is\l_\pm u_\pm^2$,
is then given implicitly by the algebraic equations
$$
\eqalign{
	&t_\pm=u_\pm^2-{1\/u_\mp}\komma\cr
	&\fun=3+{1\/u_+u_-}-2\,(u_+\+u_-)\punkt\cr}\eqn
$$
We have not yet been able to eliminate $u_\pm$ from these equations (note that
solving for $u_\pm$ in terms of $t_\pm$ amounts to solving a fifth-order
equation). There is a ``miraculous identity'' analogous to the identity
$\det u\is1$ for $p\is3$, namely $u_0u_+u_-\is1$ (here, however, 
$\det u\equiv u_0^{\phantom{2}}u_+^2u_-^2$). 
The occurrence of these properties of
the solutions is intriguing.

We would also like to generalize
to non-abelian field-strengths. It is not clear how to proceed in that
case --- the simple matrix identities we have utilized in the present
note do not carry over.

A case of special interest is the 5-brane in eleven dimensions.
Also in this case the methods we have presented here need modification,
due to the presence of a 3-form field-strength. The analogue of a DBI
action is unknown for the 5-brane --- it would require a generalization
of the standard string $\b$-function calculation to open membranes.

Another urgent issue is supersymmetrization.
One might think that the form of the lagrangian (\funnallagrangian) would
not present any advantages as compared to the \DBI\ action. However, one
important thing happens that might help. When one looks for a $\k$-symmetric
action, one needs a projection operator that splits a spinor in two
equal parts --- $\k$-symmetry is one half spinor worth of fermionic
gauge symmetry, essential to reduce the number of physical fermionic degrees
of freedom to the correct one. The solution we obtained for $p\is3$
satisfied $\det u=1$ (for reasons we do not yet fully understand), which
means that we may write down such a projection operator as 
$P=\half(1\+\Gamma)$, where
$$
\Gamma={1\/24\sqrt{-\g}}\,\e^{klmn}{\Pi_k}^\k{\Pi_l}^\l{\Pi_m}^\mu{\Pi_n}^\nu
		\Gamma_{\k\l\mu\nu}\komma\eqn
$$
where now $g_{mn}={\Pi_m}^\mu\Pi_{n\mu}$.
The fact that $\det\g\=\det g$ ensures that $\Gamma^2\=1$. This observation
makes the hope of finding a supersymmetric D-brane action for $p\is3$
less far-fetched, even if the non-linearities may become difficult to
deal with.

Finally, one should investigate the relevance of the case $\D\is0$ 
to F-theory. It is striking that we have a very simple action for the
case of selfdual field-strength. Although the constraint structure
cannot be believed to be complete, one can hope to get some hints
from this formulation, especially after supersymmetrization.

\refout
\end